\documentstyle[12pt,epsfig]{article}
\def\bge{\begin{equation}}
\def\ene{\end{equation}}
\def\bg{\begin{eqnarray}}
\def\en{\end{eqnarray}}
\def\nn{\nonumber}

\def\S0{{\Sigma^0}}

\def\k0bar{\bar{K}^0}

\begin{document}
%%%%%%%%%%%%%%%%%%%%%%%%%%%%%%%%%%%%%%%%%%%%%%%%%%%%%%%%%%%%%%%%%%%%%%%%
\renewcommand{\thefootnote}{\fnsymbol{footnote}}
\begin{flushright}
ADP-99-6/T351
\end{flushright}
\begin{center}
{\LARGE A quark-meson coupling model with short-range quark-quark 
correlations}
\end{center}
\vspace{0.5cm}
\begin{center}
%\begin{large}
K.~Saito\footnote{ksaito@nucl.phys.tohoku.ac.jp} \\
Physics Division, Tohoku College of Pharmacy \\
Sendai 981-8558, Japan \\
K.~Tsushima\footnote{ktsushim@physics.adelaide.edu.au} and 
A.W.~Thomas\footnote{athomas@physics.adelaide.edu.au} \\
Special Research Center for the Subatomic Structure of Matter \\
and Department of Physics and Mathematical Physics \\
The University of Adelaide, Adelaide, SA 5005, Australia 
%\end{large}
\end{center}
\vspace{0.5cm}
\begin{abstract}
Short-range quark-quark correlations are introduced into the quark-meson 
coupling (QMC) model in a simple way.  The effect of these correlations 
on the structure of the nucleon in dense nuclear matter is studied. 
We find that the short-range correlations may serve to reduce a 
serious problem associated with the modified quark-meson coupling 
model (within which
the bag constant is allowed to decrease with increasing density), namely the 
tendency for the size of the bound nucleon to increase rapidly as the density
rises. We also find that, with the addition of correlations, both QMC 
and modified QMC are consistent with the phenomenological equation of state 
at high density. 
\end{abstract}
%
%%%%%%%%%%%%%%%%%%%%%%%%%%%%%%%%%%%%%%%%%%%%%%%%%%
\newpage

About a decade ago, Guichon~\cite{guichon} proposed a relativistic 
quark model for nuclear matter, where it consists of non-overlapping 
nucleon bags bound by the self-consistent exchange of scalar ($\sigma$) 
and vector ($\omega$) mesons in mean-field approximation (MFA).  
This model has been further developed as the quark-meson 
coupling (QMC) model, and applied to various phenomena in nuclear physics 
(for recent reviews, see Ref.\cite{review}).  Recently, Jin and 
Jennings~\cite{jin} have proposed an alternative version of QMC (called the 
modified QMC, MQMC), where the bag constant is allowed to decrease as a 
function of density. 

So far, the use of the QMC model has been limited to the region of small to 
moderate densities, because it has been assumed that the 
nucleon bags do not overlap.  It is therefore of great 
interest to explore ways to extend the model to include 
short-range quark-quark correlations, which 
may occur when nucleon bags overlap at high density.  
In this paper we will introduce these short-range correlations in 
a very simple way, and calculate their effect on the quark 
structure of the nucleon in medium.  We refer to this model as 
the quark-meson coupling (or modified quark-meson coupling) model with 
short-range correlations (QMCs (or MQMCs)).  

Let us consider uniformly distributed (iso-symmetric) 
nuclear matter with density $\rho_B$. 
At high density the nucleon bags start to overlap with each other, and 
a quark in one nucleon may interact with quarks in other nucleons in the 
overlapping region.  
Since the interaction between the quarks is 
short range, it seems reasonable to treat it in terms of contact interactions. 
An additional interaction term of the form, 
${\cal L}_{int} \sim \sum_{i \neq j}{\bar \psi}_q(i) \Gamma_\alpha
\psi_q(i) {\bar \psi}_q(j) \Gamma^\alpha \psi_q(j)$, 
may then be added to the original QMC Lagrangian density~\cite{review}. 
Here $\psi_q(i)$ is a quark field in the $i$-th nucleon and 
$\Gamma_\alpha$ stands for $1, \gamma_5, \gamma_\mu, 
\gamma_5\gamma_\mu$ or $\sigma_{\mu\nu}$ (with or without
the isospin and color generators). 
(For the present we consider only u and d quarks.)  
In MFA most of these terms vanish in 
a static, spin saturated, uniform system because of rotational 
symmetry, parity etc.  
We shall retain only the dominant MFA contributions, namely the scalar- and 
vector-type interactions: $\Gamma_\alpha = 1$ and $\gamma_0$. 

Next we consider the probability for the nucleon bags to overlap, using a 
simple geometrical approach.  Let us first consider a collection of  
rigid balls, with a radius $R_c$.  In the close-packed structure 
of nuclear matter we find that 
the effective volume per ball is given by $V_c = 4\sqrt{2} R_c^3$.  
The corresponding density, $\rho_c$, is then given by the inverse of $V_c$, 
and hence the radius of the rigid ball is related to the density as 
$R_c = 1/(4\sqrt{2}\rho_c)^{1/3}$.  Returning to our problem, we see that  
for a given nuclear density $\rho_B$, if the nucleon bag radius, $R$, 
which is given by solving the nuclear matter problem self-consistently, 
is larger than $R_c (= 1/(4\sqrt{2}\rho_B)^{1/3})$ the nucleons will overlap. 
If $R \leq R_c$, there is no overlap. Of course, one could build a more 
sophisticated model, allowing for nucleon motion and nucleon-nucleon 
correlations~\cite{detar}, but we believe that the present model is 
sufficient for an initial investigation.			

Now consider two nucleon bags separated by a distance $d$ in nuclear 
matter. They will overlap for $d < 2R$ and the  
common volume is then given by 
$V_{ov} = V_N ( 1 - 3y/4 + y^3/16)$~\cite{resc}, where 
$V_N$ is the nucleon volume ($=4\pi R^3/3$) and $y = d/R$. It is  
natural to choose the probability of overlap, $p$, 
to be proportional to $V_{ov}/V_N$:  
\bge
p(y) \propto 1 - \frac{3y}{4} + \frac{y^3}{16}.  \label{prob} 
\ene
Of course, this choice is quite model-dependent.  In principle we could  
use an arbitrary, smooth function, which goes to unity at $y=0$ and 
zero beyond $y=2$ and which respects the three dimensional geometry of 
this problem.  
In this exploratory study, we take this simple form as an example.  

In mean-field approximation the Dirac equation for a quark field 
in a nucleon bag is given by 
\bge
[i\gamma\cdot\partial - (m_q - g_\sigma^q\sigma + 
f_s^q \langle {\bar \psi}_q \psi_q \rangle) - (g_\omega^q\omega + 
f_v^q \langle \psi_q^\dagger \psi_q \rangle)\gamma_0] \psi_q = 0, 
\label{dirac}
\ene
where $m_q$ is the bare quark mass, 
$\sigma$ and $\omega$ are the mean-field values of the $\sigma$ and 
$\omega$ mesons and $g_\sigma^q$ and $g_\omega^q$ are, respectively,  
the $\sigma$- and $\omega$-quark coupling constants in the 
usual QMC model~\cite{review}.  The new coupling constants, 
$f_{s(v)}^q$, have been introduced
for the scalar~(vector)-type short-range correlations, 
and are given by (see Eq.(\ref{prob})) 
\bge
f_{s(v)}^q = \frac{{\bar f}_{s(v)}^q}{M^2} \times (1 - \frac{3y}{4} 
+ \frac{y^3}{16}) \theta(y) \theta(2-y). 
\label{coup}
\ene
We have also taken $y = d/R$, with $d$ the   
average distance between two neighbouring nucleons at a given nuclear density 
$\rho_B$ -- i.e., as explained above,    
$d = 2R_c = 2/(4\sqrt{2}\rho_B)^{1/3}$.  Note 
that since the coupling constants have dimension of (energy)$^{-2}$  
we introduce new, dimensionless coupling constants, ${\bar f}_s^q$ and 
${\bar f}_v^q$ by dividing by the free nucleon mass ($M = 939$ MeV) squared. 
(If the coupling strength is positive the 
correlation gives a repulsive force.)  
In Eq.(\ref{dirac}), $\langle {\bar \psi}_q \psi_q \rangle$ and 
$\langle \psi_q^\dagger \psi_q \rangle$ are, respectively, the average 
values of the quark scalar density and quark density with respect to 
the nuclear ground state, which 
are approximately given by the values at the center of the nucleon   
in local density approximation~\cite{blun} 
(we will revisit this later).  

Now we can solve the Dirac equation, Eq.(\ref{dirac}), as in the usual 
QMC, with the effective quark mass 
\bge
m_q^\star = m_q - g_\sigma^q\sigma + 
f_s^q \langle {\bar \psi}_q \psi_q \rangle, 
\label{mstar}
\ene
instead of the bare quark mass.  The Lorentz vector interaction shifts the 
nucleon energy in the medium~\cite{qmc}: 
\bge
\epsilon({\vec k}) = \sqrt{M^{\star 2} 
+ {\vec k}^2} + 3(g_\omega^q \omega + f_v^q 
\langle \psi_q^\dagger \psi_q \rangle),  
\label{nenergy}
\ene
where $M^{\star}$ is the effective nucleon mass, which is given by 
the usual bag energy 
\bge
M^{\star} = \frac{3\Omega - z}{R} + \frac{4}{3}\pi BR^3.  
\label{bag}
\ene
Here $B$ and $z$ are respectively the bag constant and usual parameter 
which accounts for zero-point motion and gluon fluctuations~\cite{qmc}.  
The quark energy, $\Omega$ (in units of $1/R$), is defined by 
$\sqrt{x^2 + (Rm_q^{\star})^2}$, 
where $x$ is the lowest eigenvalue of the quark, which is given by the 
usual boundary condition at the bag surface~\cite{qmc}.  

The total energy per nucleon at density $\rho_B$ is then 
expressed as 
\bg
E_{tot} &=& \frac{4}{(2\pi)^3\rho_B} \int^{k_F} d{\vec k}  
\sqrt{M^{\star 2}+ {\vec k}^2} +  3(g_\omega^q \omega + f_v^q
\langle \psi_q^\dagger \psi_q \rangle) \nn \\ 
&+& \frac{1}{2\rho_B} (m_\sigma^2 \sigma^2 - m_\omega^2 \omega^2), 
\label{etot}
\en
where $k_F$ is the Fermi momentum, and $m_\sigma$ and $m_\omega$ are 
respectively the $\sigma$ and $\omega$ meson masses.  
The $\omega$ field created by the uniformly distributed nucleons is 
determined by baryon number conservation: 
$\omega = 3g_\omega^q\rho_B / m_\omega^2 = 
g_\omega\rho_B / m_\omega^2$ (where $g_\omega = 3g_\omega^q$), while   
the $\sigma$ field is given by the thermodynamic condition: 
$(\partial E_{tot}/\partial \sigma) = 0$.  
This gives the self-consistency 
condition (SCC) for the $\sigma$ field~\cite{qmc}: 
\bge
\sigma = - \frac{4}{(2\pi)^3 m_\sigma^2} \left( \frac{\partial M^\star} 
{\partial \sigma} \right) \int^{k_F} d{\vec k} \frac{M^\star}
{\sqrt{M^{\star 2}+ {\vec k}^2}}, 
\label{scc}
\ene
where 
\bge
\left( \frac{\partial M^\star}{\partial \sigma} \right) 
 = -3 g_\sigma^q S_N(\sigma) = - g_\sigma C_N(\sigma).  
\label{deriv}
\ene
Here $g_\sigma = 3g_\sigma^q S_N(0)$ and $C_N(\sigma) = 
S_N(\sigma)/S_N(0)$, with the quark scalar charge defined by $S_N(\sigma) = 
\int_{bag} d{\vec r} \ {\bar \psi}_q \psi_q$.  
We should note that because the scalar-type correlation does not directly 
involve the $\sigma$ field the SCC is not modified by it.  
However, the correlations do affect the $\sigma$ field through the quark 
wave function.  

In actual calculations, the quark density, 
$\langle \psi_q^\dagger \psi_q \rangle$, in the total energy, 
Eq.(\ref{etot}), may be replaced by $3\rho_B$, and the quark 
scalar density, contributing to the effective quark mass, $m_q^\star$, is 
approximately given as   
$\langle {\bar \psi}_q \psi_q \rangle = (m_\sigma^2/g_\sigma) \sigma$ 
because of the SCC (see also Ref.\cite{blun}).  

Now we present the numerical results.  First, we choose $m_q$ = 5 MeV 
and the bag radius of the nucleon in free space, $R_0$, to be 0.8 fm.  
We calculate the matter propereties using not only QMC but also 
MQMC.  For the latter we take a simple variation of the bag constant in the 
medium to illustrate the role of the short-range correlations: 
$(B/B_0)^{1/4} = \exp (-g_\sigma^B \sigma /M)$ with $g_\sigma^B$ = 
2.8~\cite{qmc1} and the bag constant in free space, $B_0$. 
In both models, the bag constant (in free space) and $z$ parameter are 
determined to fit the free nucleon mass with $R_0$ = 0.8 fm. We find 
$B_0^{1/4}$ = 170.0 MeV (in QMC, $B=B_0$ at all densities) and $z$ = 3.295.  
The coupling constants, $g_\sigma$ and $g_\omega$, are determined so as 
to reproduce the binding energy ($-15.7$ MeV) at the saturation density 
($\rho_0$ = 0.15 fm$^{-3}$).  We find that $g_\sigma^2$ = 67.80 and 
$g_\omega^2$ = 66.71 for QMC and $g_\sigma^2$ = 35.69 and
$g_\omega^2$ = 80.68 for MQMC.  Note that the matter properties at 
$\rho_0$ in both models with short-range correlations  
are identical to those of the original models~\cite{qmc,qmc1}.  
This is because, in our simple geometric approach, the effect of nucleon
overlap starts beyond $\rho_0$ (see below).  

\begin{figure}[thb]
\begin{center}
\epsfig{file=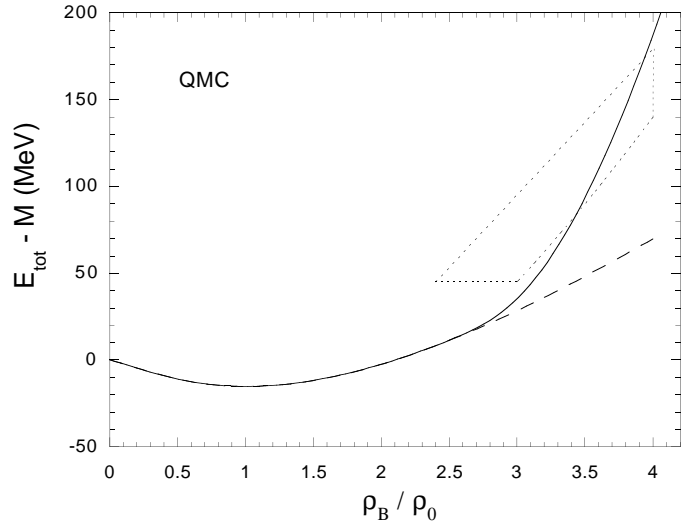,height=7cm}
\caption{Energy per nucleon for symmetric nuclear matter.  
The dashed curve is the result of the original QMC.  The solid 
curve is for the QMC with short-range correlations (QMCs).  
The region enclosed with the dotted curves 
is the empirical equation of state~\protect\cite{sano}.  
 }
\label{f:eos1}
\end{center}
\end{figure}
\begin{figure}[bht]
\begin{center}
\epsfig{file=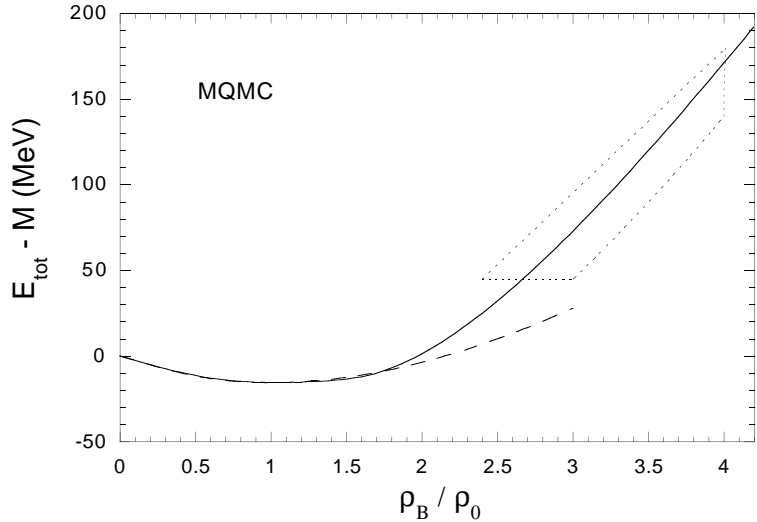,height=7cm}
\caption{
Same as in Fig.\protect\ref{f:eos1} but for MQMC.  
The dashed curve is the result of the original MQMC.  The solid
curve is for the MQMC with short-range correlations (MQMCs).  
(In the original MQMC the solution of the quark eigenvalue beyond
$\rho_B/\rho_0 \sim$ 3 cannot be found~\protect\cite{muller}.)
 }
\label{f:eos2}
\end{center}
\end{figure}
In Figs.\ref{f:eos1} and \ref{f:eos2}, we present the
total energies per nucleon for QMC and MQMC, respectively.  We determine the
coupling constant, ${\bar f}_v^q$, so as to reproduce the empirical
value of the energy around $\rho_B/\rho_0 = 2.5 \sim 4$~\cite{sano}.  
This yields the value ${\bar f}_v^q$ = 300 for QMCs and 
${\bar f}_v^q$ = 10 for MQMCs.  
(Note that in MQMCs the overlap probability is much larger than that 
in QMCs at the same density because the bag radius in MQMC increases 
very rapidly at finite density.  This is the reason why the strength of 
${\bar f}_v^q$ for MQMCs is much weaker than that in QMCs.)  
For the strength of the scalar-type correlation, since we have no
definite guideline, we take the same value in both models:
${\bar f}_s^q$ = 200 (as an illustration).  From the figures 
we can see that the nucleon overlap starts around 
$\rho_B/\rho_0 \sim$ 2.7 or 1.3 for QMCs and MQMCs, respectively.  
The empirical energies at high densities (the region enclosed with the 
dotted curves in Figs. 1 and 2~\cite{sano}) are 
well reproduced in both models (in particular, in MQMCs) if the 
short-range correlations are considered.  

\begin{figure}[thb]
\begin{center}
\epsfig{file=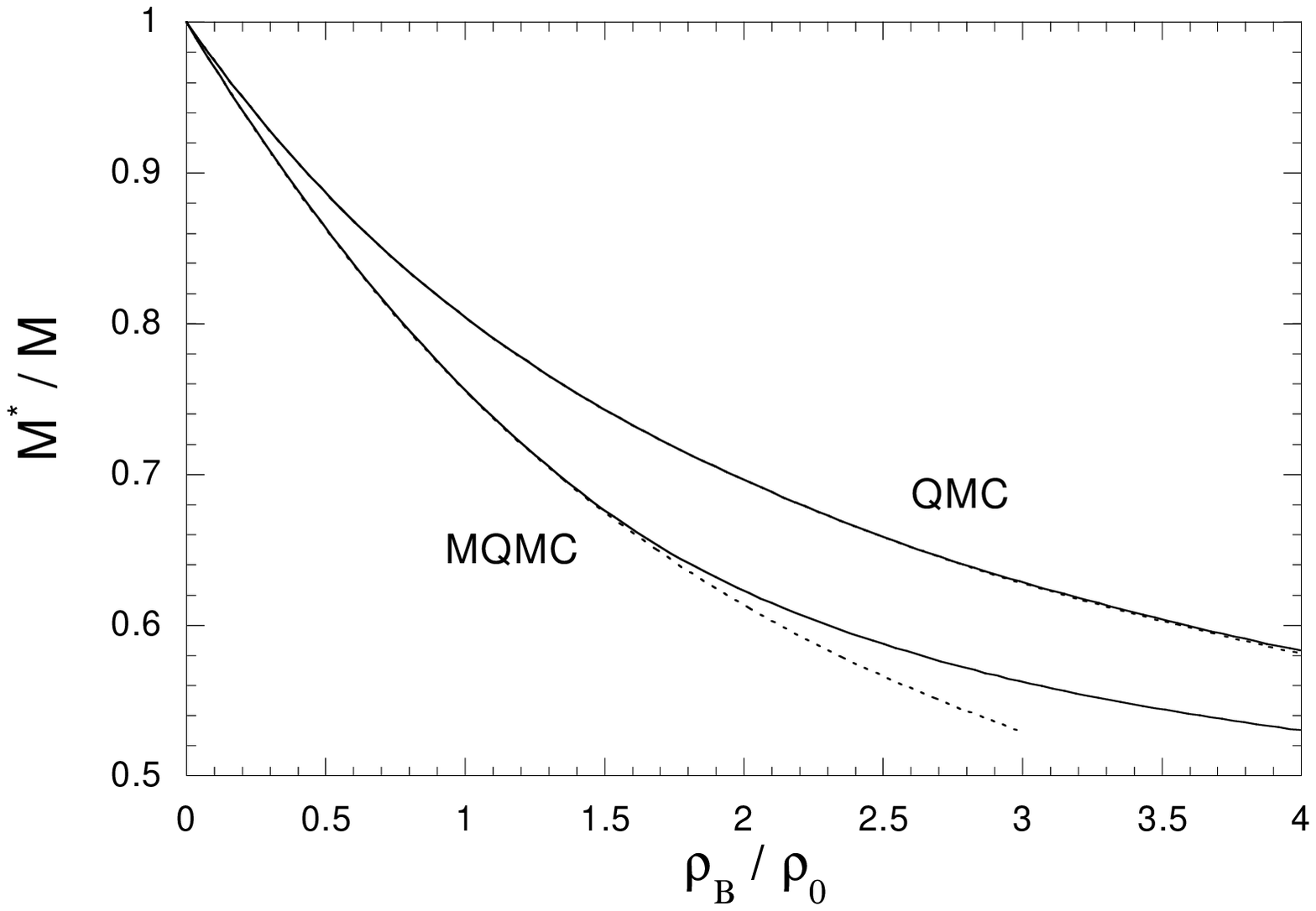,height=7cm}
\caption{
Ratio of the effective nucleon mass to the free mass.  
The dotted curves are the results of the original QMC and MQMC.  The solid
curves are for MQMCs and MQMCs. 
 }
\label{f:mass}
\end{center}
\end{figure}
\begin{figure}[bht]
\begin{center}
\epsfig{file=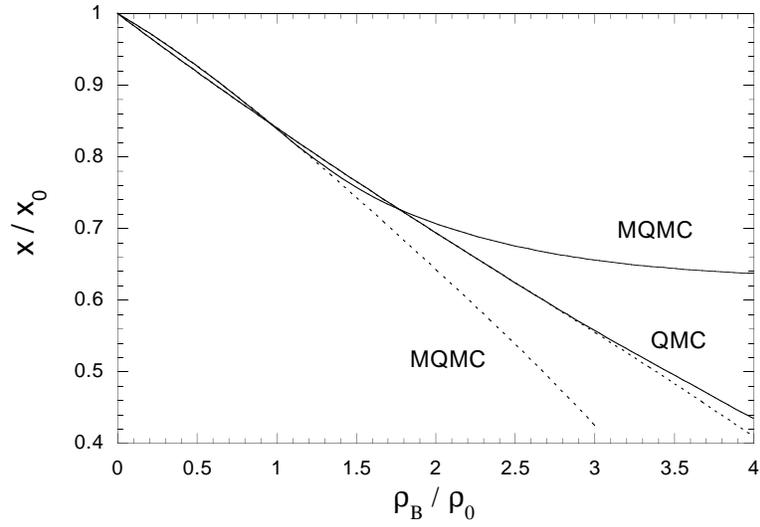,height=7cm}
\caption{
Ratio of the (lowest) quark eigenvalue in the matter to that in free 
space ($x_0$ = 2.052).  
The dotted curves are the results of the original QMC and MQMC.  
The solid curves are for QMCs and MQMCs. 
 }
\label{f:xxx}
\end{center}
\end{figure}
\begin{figure}[thb]
\begin{center}
\epsfig{file=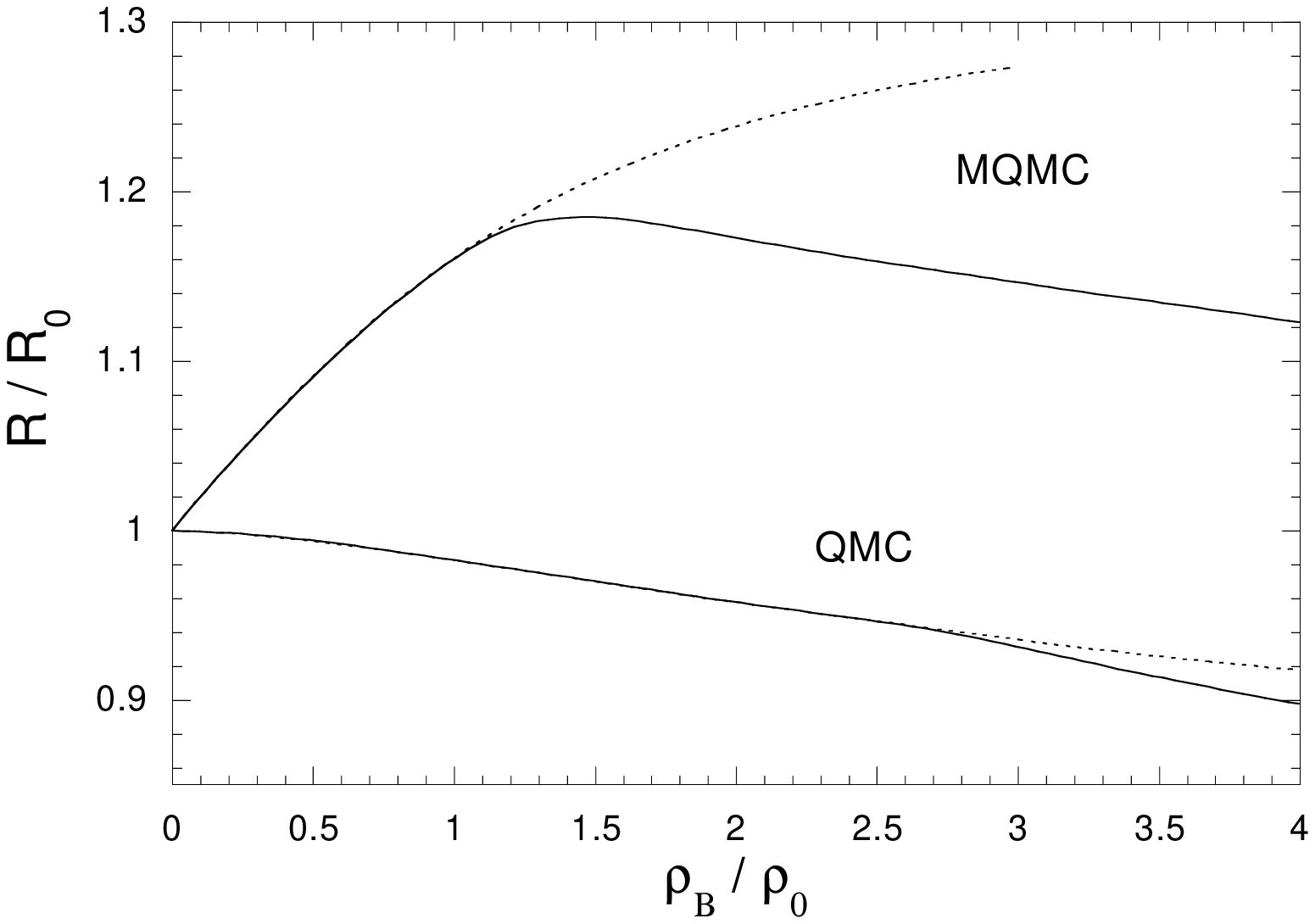,height=7cm}
\caption{
Ratio of the in-medium bag radius of the nucleon to that in free
space ($R_0$ = 0.8 fm). 
The dotted curves are the results of the original QMC and MQMC.
The solid curves are for QMCs and MQMCs.
 }
\label{f:rrr}
\end{center}
\end{figure}
In Fig.\ref{f:mass}, we show the change of the nucleon mass in matter.
We can see that the effect of the short-range correlations on the mass
is not strong.
Correspondingly, the strength of the $\sigma$ field in matter is
not much altered by the correlations.  

In Fig.\ref{f:xxx}, we present the variation of the quark eigenvalue. 
In QMCs, the effect of the short-range correlations is weak, while in 
MQMCs the effect becomes very large as the nuclear density increases and 
the nucleons overlap more and more.  Since we chose a repulsive 
scalar-type correlation, the effective 
quark mass in QMCs or MQMCs becomes larger than that in the original 
model as the density grows. This leads 
to a larger eigenvalue in QMCs or MQMCs.  As a consequence of this repulsive 
correlation a solution for the quark eigenvalue in MQMCs can be found 
even beyond $\rho_B/\rho_0 \sim$ 3.  

Turning next to the size of the nucleon itself, as measured by 
the bag radius, we see in Fig.\ref{f:rrr} that the effect of the 
short-range correlations can be very significant.  
While the effect is small in QMCs, 
in MQMCs the bag radius starts to shrink as soon as the nucleons begin to
overlap. We find a similar effect on 
the root mean square radius of the quark wave function.   
In the original MQMC 
it is well known that there is a serious problem concerning  
the bag radius. In particular, it grows rapidly at high 
density~\cite{jin,qmc1,muller} because of the decrease of the bag constant.  
However, as we can see from the figure,  
the inclusion of a repulsive (scalar-type) short-range 
correlation yields a remarkable improvement for the in-medium nucleon 
size in MQMC.  

In summary, we have studied (in mean-field approximation) the effect of 
short-range quark-quark correlations associated with nucleon overlap.  
We have found that the empirical equation of state
at high density can be very well reproduced using a repulsive vector-type 
correlation.  Furthermore, we have shown that a repulsive 
scalar-type  
correlation can counteract the tendency for the in-medium nucleon size 
to increase in MQMC. This may prove to be a significant improvement
because there are fairly strong experimental constraints on the possible 
increase in nucleon size in-medium~\cite{qmc1}.  
While our inclusion of correlations
has been based on quite simple, geometrical considerations, in the future 
we would hope to formulate the problem  
in a more sophisticated, dynamical way~\cite{mahaux} 
and to use it to study the 
properties of finite nuclei (including hyper nuclei~\cite{hyper}).  

\vspace{1cm}
This work was supported by the Australian Research Council and 
the Japan Society for the Promotion of Science.  
\clearpage
%
%%%%%%%%%%%%% Bibliography %%%%%%%%%%%%%%%%%%%%%%%%%%%%%%%%%
%

%
\end{document}